\documentclass[a4paper,11pt]{article}
\usepackage{aaskaiid}
\usepackage{orcidlink}

\usepackage{makecell}
\usepackage{threeparttable}
\usepackage{comment}
\usepackage{enumitem}
\usepackage[normalem]{ulem}

\title{Enhancing VLBI Capability with the SKA-Mid and the Jingdong 120-m Radio Telescope}
\ShortTitle{SKA-VLBI with JRT}

\author[1,2]{Wen Chen\orcidlink{0000-0002-5519-0628}}
\ShortName{Wen Chen et al.} 
\author[3]{Jun Yang\orcidlink{0000-0002-2322-5232}}
\author[1,2]{Zhixuan Li\orcidlink{0000-0001-8021-4559}}
\author[4]{Yingjie Li\orcidlink{0000-0001-7526-0120}}
\author[5, 6]{Niu Liu\orcidlink{0000-0003-2778-002X}}

\affiliation[1]{Yunnan Observatories, Chinese Academy of Sciences, Yangfangwang 396th, Kunming 650011, China}
\emailAdd{chenwen@ynao.ac.cn}
\affiliation[2]{State Key Laboratory of Radio Astronomy and Technology, Yunnan Observatories, Chinese Academy of Sciences, Kunming 650011, China}
\affiliation[3]{Department of Space, Earth and Environment, Chalmers University of Technology, Onsala Space Observatory, SE-439 92 Onsala, Sweden}
\emailAdd{jun.yang@chalmers.se}
\affiliation[4]{Purple Mountain Observatory, Chinese Academy of Sciences, No.10 Yuanhua Road, Qixia District, Nanjing 210023, China}
\affiliation[5]{School of Astronomy and Space Science, Nanjing University, Nanjing 210023, China}
\affiliation[6]{Key Laboratory of Modern Astronomy and Astrophysics (Ministry of Education), Nanjing University, Nanjing 210023, China}

\abstract{The Jingdong Radio Telescope (JRT) is a 120-meter fully steerable radio telescope currently under construction in Jingdong County, Yunnan Province, China. Located at a relatively low latitude ($24.5^{\circ}$N), the JRT will enable observations of nearly 90\% of the sky. Equipped with two broadband single-pixel receivers covering 1–8 GHz and 6–18 GHz, and a powerful digital backend, the telescope will support single-dish studies of various radio sources---particularly millisecond pulsars for enhancing the detection of nanohertz gravitational waves. In addition to single-dish capabilities, the JRT is expected to contribute approximately 800 hours annually to international Very Long Baseline Interferometry (VLBI) observations via a standard VLBI backend. When operating in conjunction with the phased-up SKA-Mid, the JRT will significantly enhance the technical and scientific capabilities of existing VLBI networks. This paper presents a comprehensive overview of the JRT’s VLBI module and explores its potential to improve joint VLBI observations with current VLBI networks. Our analysis suggests that coordinated VLBI observations involving both the SKA-Mid and the JRT have the potential to significantly advance the field. For early sciences, we also highlight a few highly promising scientific cases, e.g. measuring the distance to PSR J0437-4715 with $<$1 ly accuracy and exploring jet formation with an event-horizon-scale resolution in M60$^{*}$.}

\begin{document}
\newcommand{\actaa}{Acta Astron.} 
\newcommand{\araa}{ARA\&A} 
\newcommand{\aar}{A\&ARv} 
\newcommand{\aapr}{A\&ARv} 
\newcommand{\ab}{Astrobiol.} 
\newcommand{\aj}{AJ} 
\newcommand{\apj}{ApJ} 
\newcommand{\apjl}{ApJL} 
\newcommand{\apjs}{ApJSS} 
\newcommand{\ao}{Appl. Opt.} 
\newcommand{\apss}{Astro. \& Space Sci.} 
\newcommand{\aap}{A\&A} 
\newcommand{\aaps}{A\&AS.} 
\newcommand{\baas}{Bull. Am. Astron. Soc.} 
\newcommand{\caa}{Chinese A\&A} 
\newcommand{\cjaa}{Chinese J. A\&A} 
\newcommand{\cqg}{Class. Quantum Gravity} 
\newcommand{\gal}{Galaxies} 
\newcommand{\gca}{Geo. Cosmo. Acta} 
\newcommand{\icarus}{Icarus} 
\newcommand{\jcap}{JCAP} 
\newcommand{\jgr}{J. Geophys. Res.} 
\newcommand{\jgrp}{J. Geophys. Res. Planets} 
\newcommand{\jqsrt}{J. Quant. Spectrosc. Radiat. Transf.} 
\newcommand{\memsai}{Mem. SAIt} 
\newcommand{\mnras}{MNRAS} 
\newcommand{\nat}{Nature} 
\newcommand{\nastro}{Nat. Astron.} 
\newcommand{\ncomms}{Nat. Commun.} 
\newcommand{\nphys}{Nat. Phys.} 
\newcommand{\na}{New Astron.} 
\newcommand{\nar}{New Astron. Rev.} 
\newcommand{\physrep}{Phys. Rep.} 
\newcommand{\pra}{Phys. Rev. A} 
\newcommand{\prb}{Phys. Rev. B} 
\newcommand{\prc}{Phys. Rev. C} 
\newcommand{\prd}{Phys. Rev. D} 
\newcommand{\pre}{Phys. Rev. E} 
\newcommand{\prx}{Phys. Rev. X} 
\newcommand{\prl}{Phys. Rev. Let.} 
\newcommand{\psj}{Planet. Sci. J.} 
\newcommand{\planss}{Planet. Space Sci.} 
\newcommand{\pnas}{Proc. Natl Acad. Sci. USA} 
\newcommand{\procspie}{Proc. SPIE} 
\newcommand{\pasa}{PASA} 
\newcommand{\pasj}{PASJ} 
\newcommand{\pasp}{PASP} 
\newcommand{\rmxaa}{RMXAA} 
\newcommand{\sci}{Science} 
\newcommand{\sciadv}{Sci. Adv.} 
\newcommand{\solphys}{Sol. Phys.} 
\newcommand{\sovast}{Soviet Ast.} 
\newcommand{\ssr}{Space Sci. Rev.} 
\newcommand{\uni}{Universe} 
\newcommand{\raa}{Res. Astron. Astrophys.} 
\setlength{\bibsep}{0.0pt} 
\maketitle

\section{Introduction}

\begin{figure}[h]
    \centering
	\includegraphics[width=1\columnwidth]{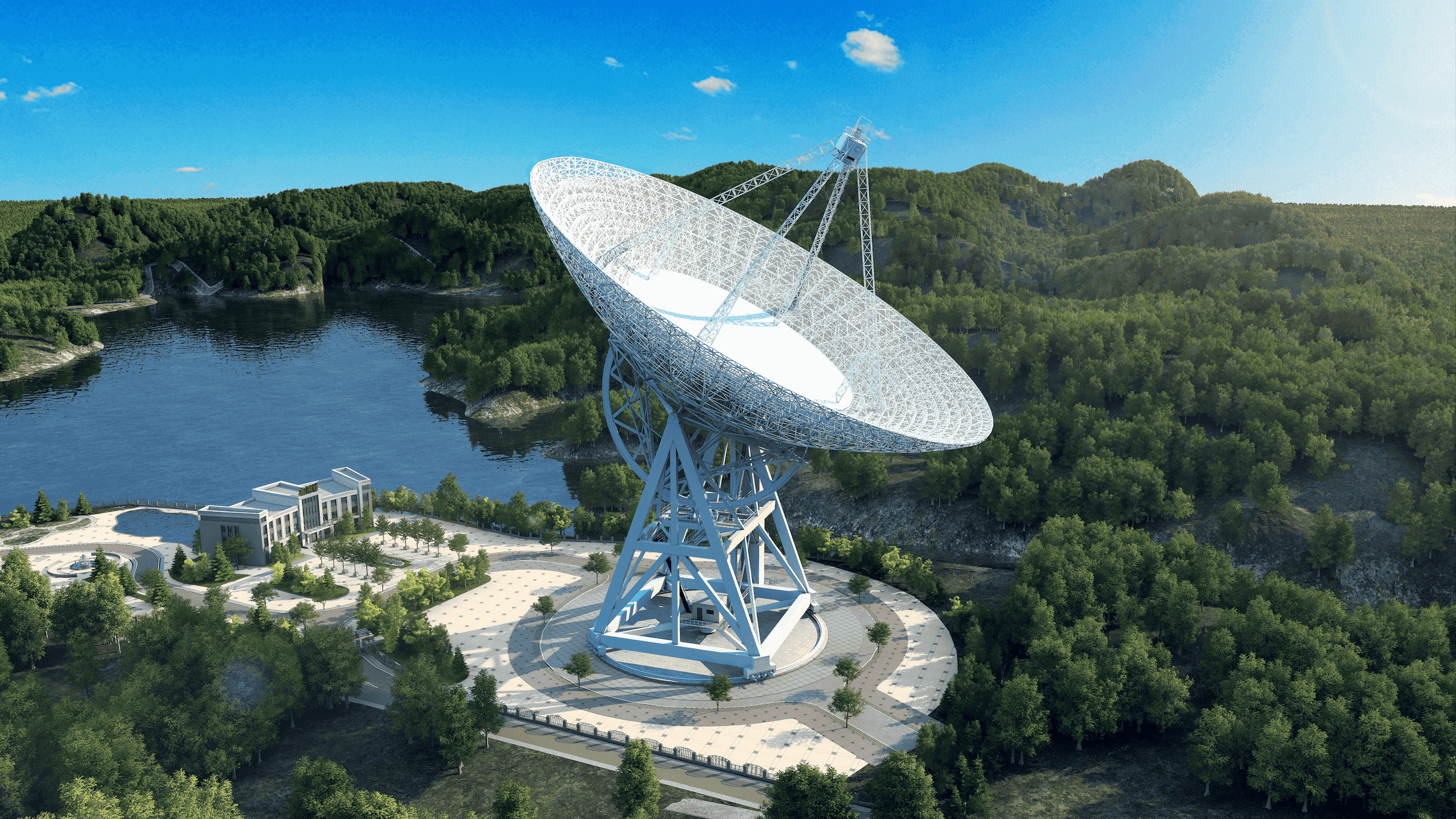}
    \caption{Concept image of the Jingdong 120-m radio telescope and the observatory layout. }
    \label{fig:telescope}
\end{figure}

Yunnan Observatories, Chinese Academy of Sciences, has conducted a systematic site survey since 2009 to find an optimal location for a next-generation large-aperture radio telescope at centimeter wavelengths. After evaluating 20 potential sites in Yunnan Province, Xujiaba in Jingdong County (Pu’er City) emerged as the best choice for this new establishment. Officially started in 2020, the project is known as the Jingdong Radio Telescope (JRT, \citealp{wangm2022}). Once finished in 2028, JRT, with a diameter of 120~m, will rank as the largest fully steerable radio telescope worldwide (see Fig.~\ref{fig:telescope}).

The main focus behind the development of JRT is pulsar science, but it is also designed with the flexibility to accommodate various other astronomical and astrophysical researches. The observing frequencies are expected to cover ranges from 0.1 to 18~GHz. Located at a relatively low latitude of $24.5^{\circ}$N, JRT provides better sky coverage than many telescopes located further north. This capability allows it to significantly complement international facilities such as the 100-m Effelsberg \citep{2011JAHH...14....3W_ef} in Germany and the Green Bank Telescope \citep{2022A&A...659A.113W} in the USA.

In addition to its role as a standalone instrument for high-precision pulsar timing, JRT will play a critical role in international Very Long Baseline Interferometry (VLBI) networks. As a large and sensitive southern station, JRT is expected to significantly enhance array sensitivity, optimize $uv$-coverage, and improve angular resolution. In particular, its synergy with SKA-Mid \citep{braun2019anticipatedperformancesquarekilometre} will enable transformational VLBI sciences on compact radio sources. This paper presents the design and specifications of JRT, evaluates its potential contributions to VLBI imaging performance, and outlines its major scientific goals.

\section{Jingdong 120-m Radio Telescope}
\subsection{The optimal location with 90\% sky accessibility}

The JRT is being constructed in Xujiaba, located in Jingdong County, Yunnan Province, China. The antenna coordinates are $24.5^{\circ}$ N and $101.0^{\circ}$ E at an altitude of approximately 1,900 meters. This location was selected after a comprehensive environmental assessment campaign that included studies of radio frequency interference (RFI), atmospheric stability examinations, and accessibility and infrastructure reviews. This area is characterized by a sparse population, very little industrial presence, and relatively stable weather conditions. Therefor, it is an optimal radio-quiet site for conducting sensitive astronomical observations.

\begin{figure}[h]
    \centering
	\includegraphics[width=1\columnwidth]{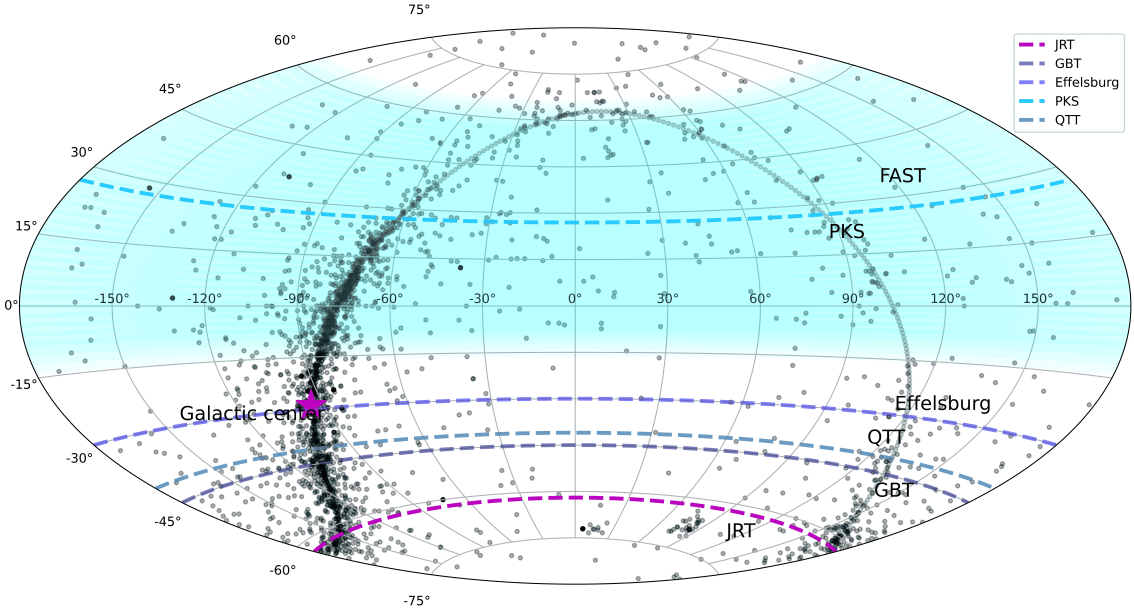}
    \caption{Known pulsars and their visibility from some large radio telescopes \citep{wangm2022}. Black dots indicate discovered pulsars.  The shaded blue area denotes the sky coverage of the Five-hundred-meter Aperture Spherical Telescope. The blue line segment marks the northern sky observation limit of the Parkes 64-m radio telescope~\citep{2020MNRAS.491.5951H}.  The dashed lines in dark blue, steel blue, navy and purple signify the observation limits of Effelsberg 100-m radio telescope~\citep{2011JAHH...14....3W_ef}, Green Bank 100-m radio Telescope (GBT, \citealp{2022A&A...659A.113W}), Qitai 110-m radio Telescope (QTT, \citealp{2014SSPMA..44..783W}), and JRT in the southern sky, respectively. The position of the center of the Milky Way is marked by a five-pointed star.}
    \label{fig:sky_cover}
\end{figure}

Fig.~\ref{fig:sky_cover} demonstrates the sky coverages of the JRT and several other major radio telescopes. Positioned at a relatively low latitude, the JRT is capable of observing celestial bodies within the declination range of approximately $-60^{\circ}$ to $+90^{\circ}$. This capacity allows the JRT to view nearly 90\% of the sky, incorporating the Galactic Center and much of the southern sky, areas not accessible to significant northern hemisphere facilities such as Effelsberg (100~m, $50^{\circ}$N, \citealp{2011JAHH...14....3W_ef}) and the Green Bank Telescope (GBT, 110~m, $38^{\circ}$N, \citealp{2022A&A...659A.113W}). The low-latitude FAST ($25^{\circ}$N, \citealp{2019SCPMA..6259502J}) has relatively small sky coverage because of its special design. In addition, as a fully steerable steerable parabolic dish, JRT facilitates continuous target tracking and extended monitoring campaigns together with other international telescopes in Europe and North America.

\subsection{The antenna design and system diagram}

JRT adopts a fully steerable design and has a primary reflector with a diameter of 120~m. The basic antenna and system parameters are listed in Table~\ref{spcifications_JRT}. The telescope is designed to operate across a broad frequency range of 100~MHz to 18~GHz, depending on the final surface accuracy. The surface panels are required to achieve a root-mean-square (RMS) accuracy better than 0.3~mm, ensuring efficient operation at frequencies up to 18~GHz. The reflector and backup truss system are designed to ensure a pointing accuracy of about 10~arcseconds under nominal wind loading conditions.

\begin{table}
	\centering
	\caption{Basic specifications for the observation system of the JRT. }
	\label{spcifications_JRT}
        \begin{threeparttable}
	\begin{tabular}{cc}
	\hline
  	 Item                   & Details                \\
    \hline
    Location                 & {$101^{\circ}$E, $24.5^{\circ}$N} \\
    Main reflector           & 120~m (inner 80 m: solid panel, outer 40 m: mesh panel)          \\
	Mount                    & AZ/EL, Wheel-rail type             \\ 
	  f/D                      & 0.42 at the primary focus          \\ 
	Polarization             & Dual-linear          \\ 
	Frequency range          & $\leq$18~GHz          \\ 
	  Panel accuracy (rms)     & $<$0.3 mm for all   \\
    System temperature       &   $\sim$25~K        \\
    SEFD                     &  $\sim$10~Jy      \\
    Pointing accuracy        &  $\leq$10$^{\prime\prime}$          \\
    Aperture efficiency      &       $\geq 60\%$              \\
    Slewing speed            & $0.8^{\circ}$/s at AZ, $0.3^{\circ}$/s at EL    \\
    Slewing range            & $\pm270^{\circ}$ at AZ, 5--89$^{\circ}$ at EL (software EL limit $\sim$10$^{\circ}$)  \\
    Weight                   & 3400 t \\  
	\hline
	\end{tabular}
    \end{threeparttable}
\end{table}

\begin{figure}[h]
    \centering
	\includegraphics[width=0.8\columnwidth]{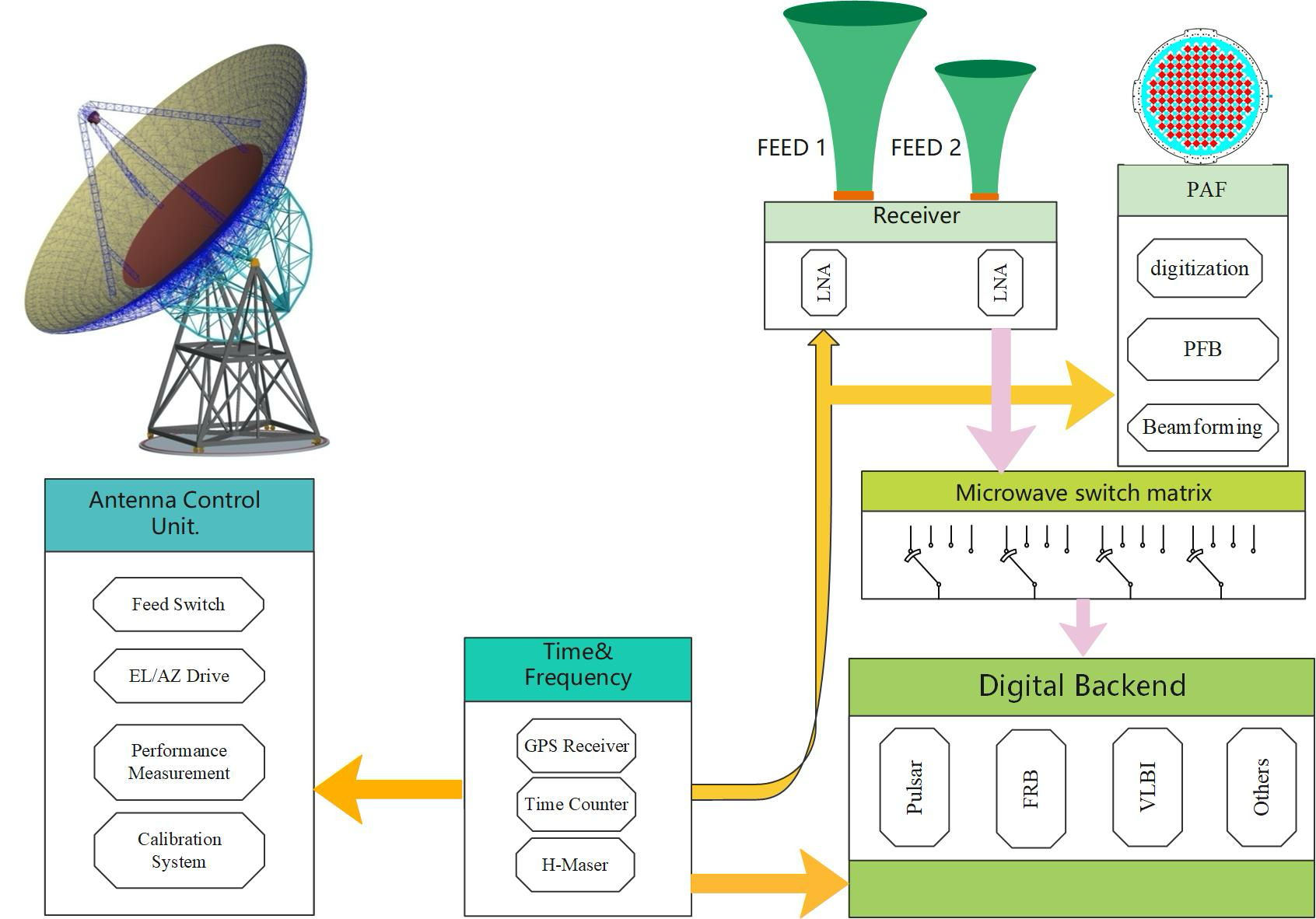}
    \caption{Overview block diagram of the JRT's multi-function observation system. Note that the PAF receiver is a future upgrade option for wide-field astronomy.  }
    \label{fig:system}
\end{figure}

The feed system employs a broadband quad-ridge horn with integrated cryogenic cooling to minimize receiver noise temperature. Currently, there are only two signle-pixel broadband receivers planned.  One will cover 1--8~GHz. The other will cover 6--18~GHz. In addition, there is a position available for the installation of a phased array feed (PAF) in the receiver cabin. The PAF would be used to further enhance survey speed and field of view (see Fig.~\ref{fig:system}). This configuration will provide continuous frequency coverage for diverse scientific applications, ranging from low-frequency pulsar search to high-frequency VLBI observations of compact extragalactic sources.

The telescope employs an alt-azimuth mount with a maximum slewing speed of 0.8$^{\circ}$/s, enabling rapid source acquisition and efficient participation in VLBI experiments that require frequent source switching.

The control and data acquisition system of JRT is being developed to ensure compatibility with international standards for pulsar timing and VLBI. The VLBI backend will support wide-bandwidth digitization and real-time recording, consistent with next-generation VLBI networks. JRT can also adopt internationally standardized VLBI terminals to ensure compatibility with global networks and enable high-speed data acquisition. The telescope will also feature a flexible backend switching architecture, allowing seamless transitions between single-dish pulsar observations, spectral line studies, and VLBI modes. 

Overall, the design philosophy of JRT emphasizes both high sensitivity and operational versatility. Its combination of large collecting area, broad frequency coverage, and modern digital backends ensures that it will be a competitive instrument in the coming decades, well aligned with the scientific goals of the SKA era.

\subsection{Key scientific goals for single-dish observations}

The primary scientific objectives of JRT are to establish a pulsar-based time standard and to enable the detection of nanohertz gravitational waves through high-precision pulsar timing \citep[e.g.][]{Xu2023FAST}. In addition, the telescope will support a broad range of research, including pulsar astrophysics \citep[e.g.][]{Xu2024, Hao2025}, tests of gravity \citep[e.g.][]{Hu2025}, studies of fast radio bursts (FRBs,~\citealp{2019A&ARv..27....4P_frb}), black holes and their surrounding environments.

While pulsar timing remains the top priority, JRT is also designed to serve the wider astronomical community. The telescope is expected to contribute no less than 800 hours annually to various domestic and international VLBI observations. The inclusion of large-aperture stations such as JRT in global VLBI arrays is essential: additional stations improve overall network sensitivity, and large dishes in particular are critical for probing faint, compact radio sources.

\section{VLBI UV-coverage and Sensitivities}

The inclusion of JRT into international VLBI arrays is expected to significantly enhance both sensitivity and imaging capability. Thanks to its large collecting area and favorable southern location, JRT will provide long baselines to existing facilities in East Asia, Europe, Africa, and Australia. In particular, the synergy between JRT and SKA-Mid in South Africa will offer highly sensitive intercontinental baselines, enabling milliarcsecond-scale studies of faint and compact radio sources.

\subsection{Common-view time with SKA-Mid}

\begin{figure}[h]
 \centering
    \includegraphics[width=0.8\linewidth]{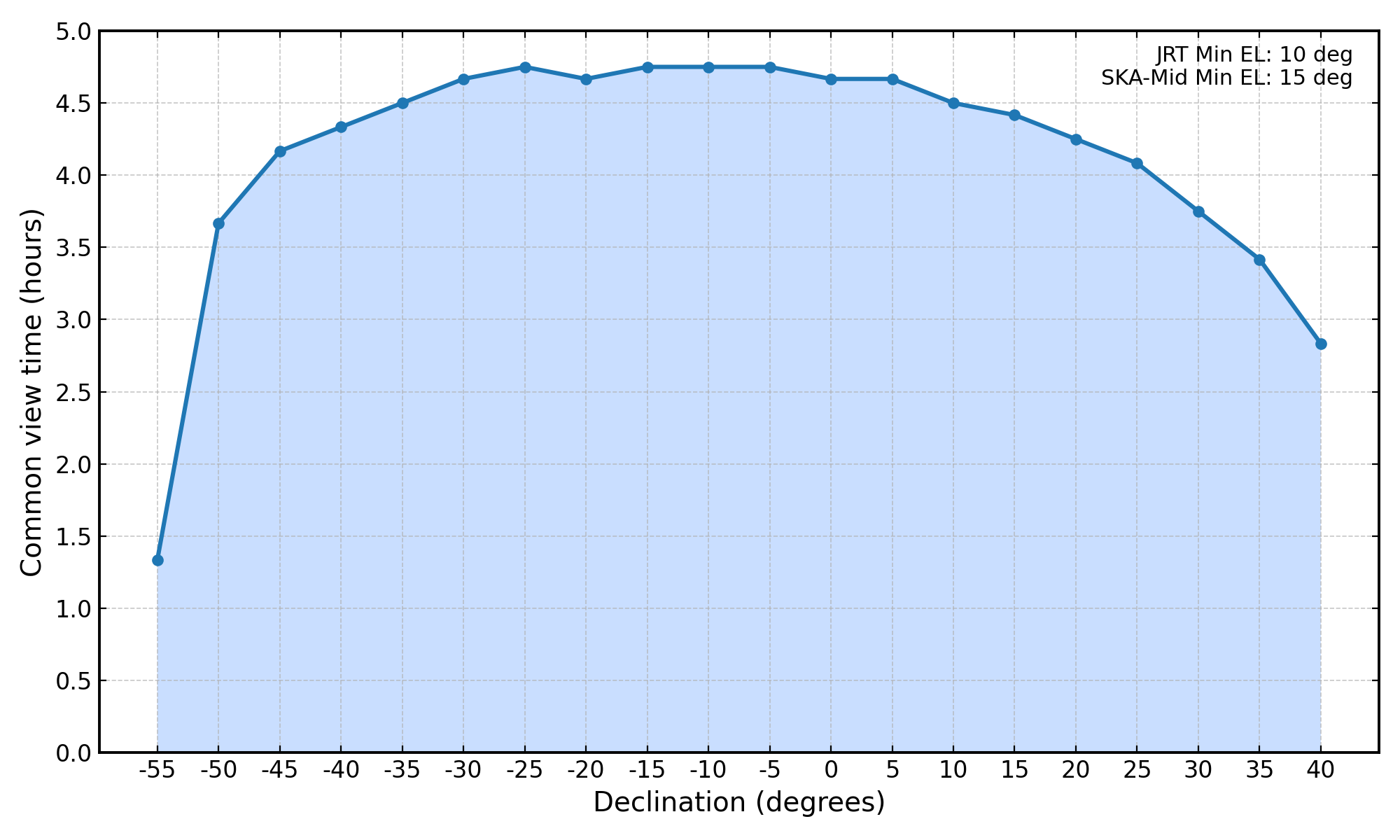}
    \caption{Common-view time for target sources at different declinations on the JRT--SKA-Mid baseline.}
\label{com_time}
\end{figure}

JRT and SKA-Mid have a substantial overlap in sky visibility. Fig.~\ref{com_time} shows the common-view time for target sources at different declinations between JRT and SKA-Mid. For sources near the celestial equator, the two telescopes can achieve more than $\sim$4~hours of common-view time per day, allowing long time observations with excellent $uv$-coverage. This configuration is particularly advantageous for astrometric programs and high dynamic range imaging.

\subsection{UV-coverage Enhancement}

\begin{figure}[htbp]
    \centering
    \includegraphics[width=0.8\linewidth]{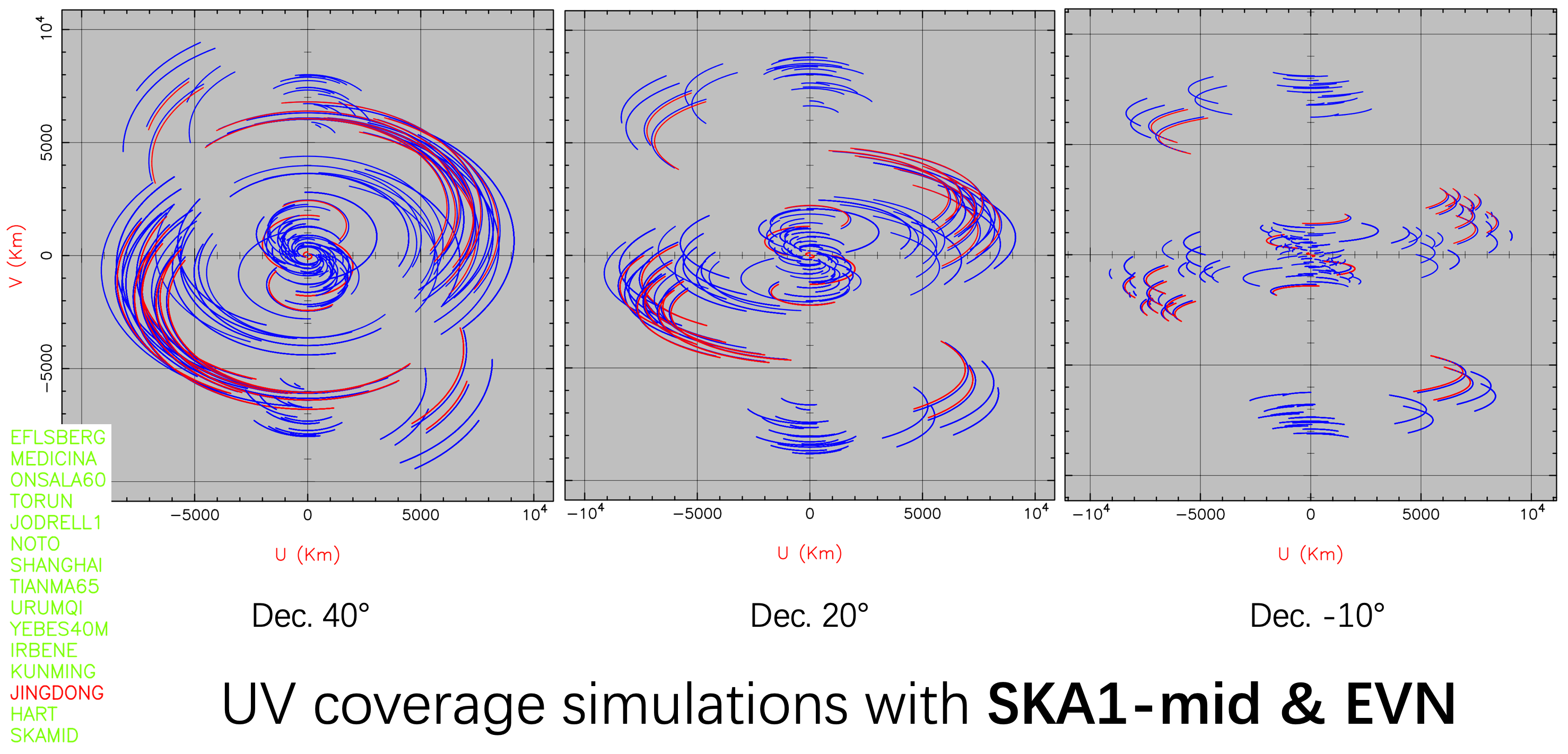}
    \includegraphics[width=0.8\linewidth]{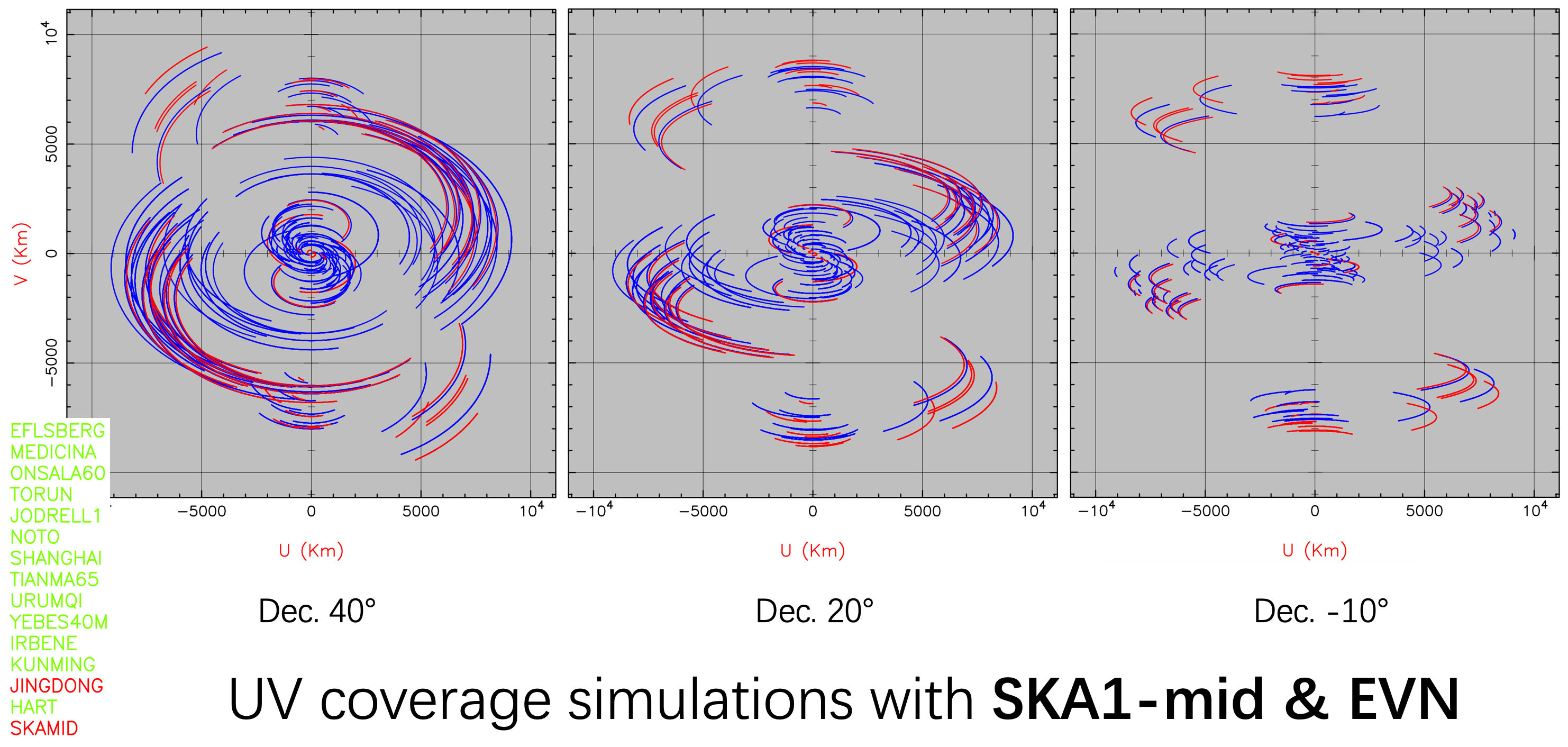}
    \caption{$uv$-coverage simulations with SKA-Mid + EVN including JRT for sources at different declinations.}
    \label{UV_SKA}
\end{figure}

\begin{figure}[htbp]
    \centering
    \includegraphics[width=0.8\linewidth]{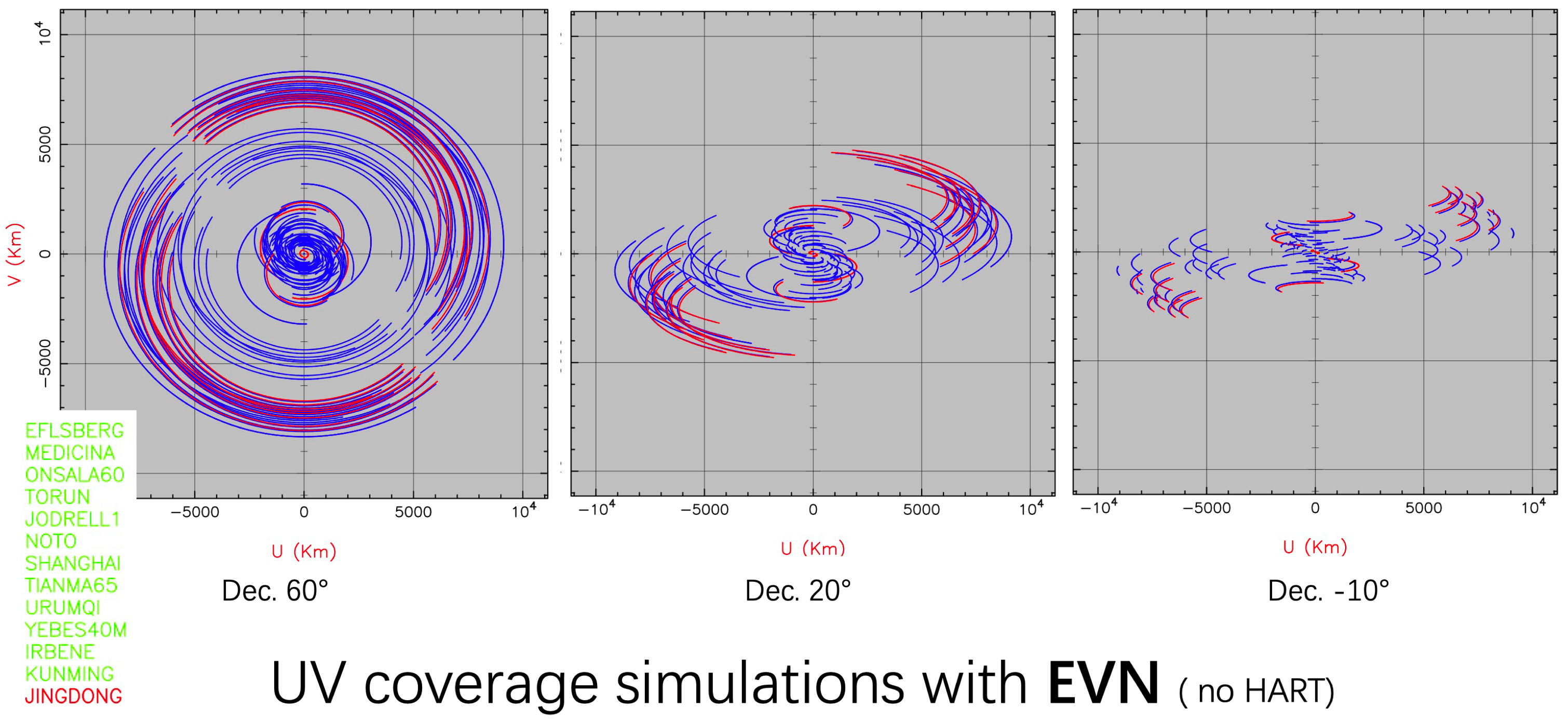}
    \caption{$uv$-coverage simulations with EVN including JRT for sources at different declinations.}
    \label{UV_EVN}
\end{figure}

\begin{figure}[htbp]
    \centering
    \includegraphics[width=0.8\linewidth]{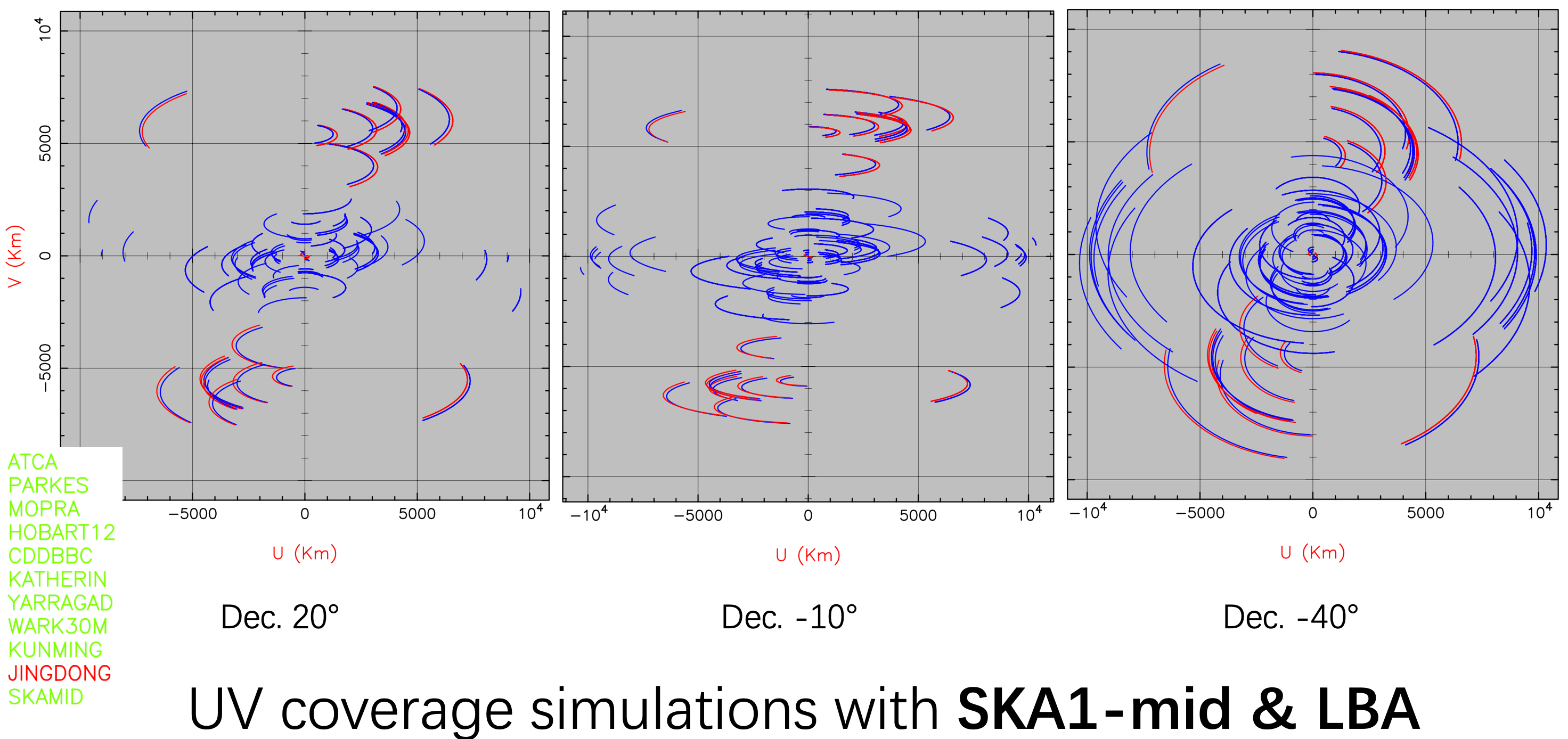}
    \includegraphics[width=0.8\linewidth]{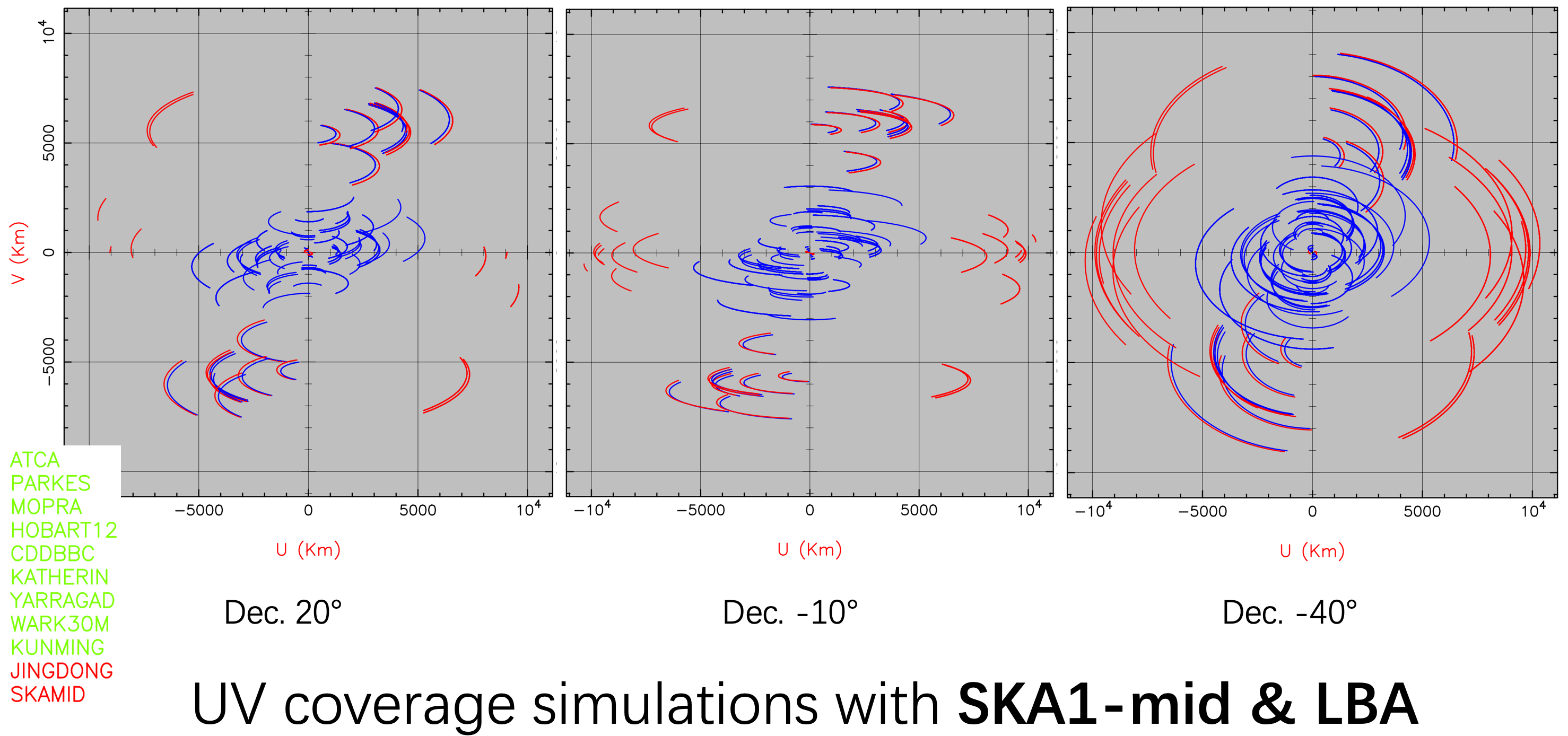}
    \caption{$uv$-coverage simulations with LBA including JRT for sources at different declinations.}
    \label{UV_LBA}
\end{figure}

Simulated $uv$-coverage plots demonstrate the impact of JRT’s inclusion in various VLBI arrays. As shown in Fig.~\ref{UV_SKA} to~\ref{UV_LBA} (the red line represents the JRT-related baseline), JRT provides long baselines to SKA-Mid and east–west extensions to European and north–south to Australian stations. This results in improved $uv$-plane sampling, which directly translates into higher image dynamic range. 

In the SKA-Mid+EVN configuration, SKA-Mid--Tianma~65-m provides the longest baselines, while SKA-Mid--JRT offers comparable baseline lengths and high-SNR measurements at the outer $uv$ coverage. The corresponding fringe detection sensitivity is summarized in Table~\ref{tab:fringe_ska_jrt_combined}.

\subsection{Image sensitivity improvements}
The improvement in image sensitivity brought by JRT has been quantified for multiple international VLBI networks. Table~\ref{sensiti} summarizes the image sensitivity assuming 1~Gbps data rate, 10~hours of observation time, and 70\% on-target efficiency. 
The improvement is substantial across all networks, ranging from a factor of 1.4 for SKA-Mid + EVN \citep{2015arXiv150105079Z_evn} to more than a factor of 3.6 for EAVN \citep{2022Galax..10..113A-eavn}.
Notably, for LBA \citep{edwardssouthern_lba}, the inclusion of JRT improves the sensitivity by more than a factor of two at both 1.7 and 6.0~GHz, enabling the detection of previously inaccessible faint sources.

\begin{table}[h]
	\centering
	\caption{VLBI network imaging performance assuming 1~Gbps data rate, 10~hours of observation time, and 70\% on-target time. }
	\label{sensiti}
        \begin{threeparttable}
	\begin{tabular}{cccccc}
	\hline
  	 VLBI   & Freq. & $\sigma_{\rm img}({\rm without~JRT})$ &  $\sigma_{\rm img}({\rm with~JRT})$ &  Beam Size& Imp. \\
    Network & (GHz) &  ($\mu$Jy/beam) &   ($\mu$Jy/beam) & w/ JRT (mas)  &  Ratio \\
        \hline
SKA-Mid \& EVN & 1.7 & 1.68 & 1.23 & 3.50 & 1.4 \\
EVN & 1.7 & 4.46 & 2.67   & 3.87  & 2.7 \\
EVN & 6.0 & 3.70 & 2.28   & 1.10 & 2.6 \\
EAVN & 6.0 & 23.0 & 6.43   & 2.45 & 3.6 \\
LBA & 1.7 & 7.74 & 3.69   & 3.35 & 2.1 \\
LBA & 6.0 & 16.6 & 6.49   & 1.00 & 2.6  \\      
	\hline
	\end{tabular}
    \begin{tablenotes}
        \footnotesize
\item 
Notes: The ``Imp.\ Ratio'' column is defined as the ratio of the image sensitivity without JRT to that with JRT, i.e., $\sigma_{\rm img}({\rm without~JRT})/\sigma_{\rm img}({\rm with~JRT})$.
    \end{tablenotes}
    \end{threeparttable}  
\end{table}


\subsection{Fringe detection sensitivity}

In VLBI, the fringe detection sensitivity on a single baseline can be estimated from the thermal noise \citep{taylor1999, HaystackMemo020}
\begin{equation}\label{eq:sigma_bl}
\sigma_{\rm bl}=\frac{1}{\eta_s\sqrt{N_{\rm pol}}}
\sqrt{\frac{{\rm SEFD}_1\,{\rm SEFD}_2}{2\,\Delta\nu\,\tau}},
\end{equation}
where ${\rm SEFD}_1$ and ${\rm SEFD}_2$ are the system-equivalent flux densities of the two stations, $\Delta\nu$ is the effective bandwidth used for fringe fitting, $\tau$ is the coherent integration time, $N_{\rm pol}$ is the number of polarizations, and $\eta_s$ is the correlator/quantization efficiency. Following the fringe-search S/N definition \citep{rogers_1995AJ....109.1391R}, we define the fringe detection limit as
\begin{equation}\label{eq:sfd_def}
S_{\rm fd}={\rm SNR}_{\rm th}\,\sigma_{\rm bl},
\end{equation}
where ${\rm SNR}_{\rm th}$ is the adopted detection threshold.

For spectral-line targets (e.g., extragalactic CH$_3$OH masers), $\Delta\nu$ should correspond to the channel width used for fringe fitting rather than the full continuum bandwidth. For a representative continuum fringe-fitting setup, we adopt $\Delta\nu=128$~MHz per polarization and $\tau=60$~s (together with $\eta_s=0.88$, $N_{\rm pol}=2$, and ${\rm SNR}_{\rm th}=5$) to evaluate $S_{\rm fd}$.

Using equation~(\ref{eq:sigma_bl}) and (\ref{eq:sfd_def}), Table~\ref{tab:fringe_ska_jrt_combined} summarizes the resulting fringe detection limits on the SKA-Mid--JRT baseline across SKA-Mid Bands~1/2/5a/5b at representative frequencies, and also provides a direct comparison to the longest SKA-Mid baselines where Tianma~65-m is available at the corresponding frequency.

Table~\ref{tab:fringe_ska_jrt_combined} shows that the SKA-Mid--JRT baseline reaches sub-mJy single-baseline detection limits across the SKA-Mid observing bands for continuum fringe fitting. This provides a lower fringe-detection threshold $S_{\rm fd}$ and improves the detectability of faint compact emission, particularly on the outer edge of the $uv$ coverage where long-baseline sensitivity is most critical. For spectral-line fringe fitting, the effective bandwidth is set by the channel width, so the thermal-noise limit scales as $S_{\rm fd}\propto(\Delta\nu\,\tau)^{-1/2}$.
For example, adopting $\Delta\nu=10$\,kHz gives $S_{\rm fd}=10.06$\,mJy at 6.7\,GHz for $\tau=180$\,s, and $S_{\rm fd}=13.98$\,mJy at 12.2\,GHz for $\tau=120$\,s on the SKA-Mid--JRT baseline.

\begin{table}[h]
\centering
\caption{Fringe detection sensitivity on the SKA-Mid--JRT baseline across SKA-Mid Bands~1,~2,~5a,~and~5b, with a comparison to SKA-Mid--TM65 at representative frequencies where available.}
\label{tab:fringe_ska_jrt_combined}
\begin{threeparttable}
\begin{tabular}{cccccc}
\hline
SKA-Mid  & Freq. range & Rep. freq. &
$S_{\rm fd}$ (SKA--TM65) & $S_{\rm fd}$ (SKA--JRT) & Imp. \\
Band & (GHz) & (GHz) & Cont./Spec.(mJy) & Cont./Spec.(mJy) & Ratio\\
\hline
1  & 0.35--1.05 & 0.8  & -- / --        & 0.163 / 9.22 & --\\
2  & 0.95--1.76 & 1.7  & 0.251 / 12.96  & 0.128 / 6.48   & 2.0 \\
5a & 4.6--8.5   & 6.7  & 0.246 / 16.08  & 0.154 / 10.06   & 1.6 \\
5b & 8.3--15.4  & 12.2 & 0.410\tnote{a} / 32.80& 0.175 / 13.98 & 2.3 \\
\hline
\end{tabular}
\begin{tablenotes}
\footnotesize
\item Notes: $S_{\rm fd}$ is the minimum correlated flux density detectable on a single baseline, computed from
Eqs.~(\ref{eq:sigma_bl})--(\ref{eq:sfd_def}) with $\eta_s=0.88$, $N_{\rm pol}=2$, and ${\rm SNR}_{\rm th}=5$.
For the continuum (Cont.) values we adopt $\Delta\nu_{\rm cont}=128$\,MHz per polarization and $\tau_{\rm cont}=60$\,s.
For the spectral-line (Spec.) values we adopt a single-channel bandwidth $\Delta\nu_{\rm spec}=10$\,kHz and coherent integration times
$\tau_{\rm spec}=240$\,s (0.8\,GHz), 300\,s (1.7\,GHz), 180\,s (6.7\,GHz), and 120\,s (12.2\,GHz).
The ``Imp. Ratio''(improvement ratio) column is defined as $S_{\rm fd}({\rm SKA\!-\!TM65})/S_{\rm fd}({\rm SKA\!-\!JRT})$.
\item[a] For TM65 at 12.2\,GHz, we adopt a representative Ku-band SEFD of 56\,Jy (typical value, \citealp{yan_2024Univ...10..195Y}).
\end{tablenotes}
\end{threeparttable}
\end{table}

\subsection{Potential VLBI Sciences}

By conducting joint VLBI observations with the JRT and the SKA, astronomers can greatly enhance both $uv$-coverage and image sensitivity. These advancements enable astronomers to uncover unprecedented details in complex jet structures, investigate jet formation and acceleration, follow the evolution of faint radio afterglows in astrophysical transients, perform high-precision astrometry on Galactic radio sources, e.g. radio stars, pulsars, and maser sources, identify the nature of compact objects (neutron stars or stellar-mass black holes) in binary systems, delve into cosmological studies, and engage in fundamental physics research. The potential of these VLBI sciences has been discussed in works by \cite{Paragi2015, Paragi2018, Li2024, XuYe01.2026.SKA, TaoAn03.2026.SKA} and \cite{Shu01.2026.SKA}. In this context, our aim is to add a few more specific cases to illustrate the significant impact.

\textbf{VLBI astrometry on PSR~J0437$-$4715 with 2-$\mu$as precision.} High-precision differential astrometry is a unique advantage for VLBI observations of nearby radio sources. Currently, VLBI astrometry on pulsars has reached a precision of $\sim$10~$\mu$as \citep[e.g.][]{Deller2019}. To strengthen the constraint on Newton's Gravitational Constant variation by time ($\dot{G}/G$) with $<$10~$\mu$as, we have studied the nearest and brightest millisecond pulsar PSR~J0437$-$4715. Our previous research revealed two faint reference sources with angular separations of $<1~^{\prime}$ and flat radio spectra \citep{Li2018}. Our 1-Gbps in-beam phase-referencing observations at 6.7~GHz have also achieved a precision of $\sim$20~$\mu$as on the parallax (Chen et al., in prep.). With the 32-Gbps data rate and the more sensitive stations, including the JRT and the SKA-Mid, we expect to improve the astrometry precision by a factor of 10 and reach an unprecedented level, $\sim$2~$\mu$as. Together with the precise kinematic distance inferred from pulsar timing observations \citep{Reardon2024}, we would provide the most stringent constraint on $\dot{G}/G$ for fundamental physics \citep{Deller2008}.

\textbf{Exploring jet formation with horizon-scale spatial resolution.} In the field of astrophysics, understanding jet formation is a significant challenge, especially in systems with extremely low accretion rates, such as Sgr A$^{*}$. The elliptical galaxy M60, located 14 Mpc away, contains a low-luminosity active galactic nucleus (AGN), designated M60$^{*}$, with an accretion rate close to 10$^{-8}$, slightly higher than Sgr A$^{*}$. It features a flat-spectrum radio core with an integrated flux density of approximately 15 mJy and a theoretical black hole shadow size of 28 $\mu$as \citep{Li2024M60}, and it produces a faint jet (Cheng et al., in prep.). Among extragalactic radio sources within a distance of less than 200 Mpc, M60$^{*}$ stands out as the only known source that closely resembles Sgr A$^{*}$ (Cheng et al., in prep.). At 8.4 GHz, observations revealed only a compact radio core with a size less than 200 $\mu$as. This straightforward structure provides a unique opportunity to thoroughly investigate the formation and emergence of transient jets on event-horizon scales during radio flares with high-sensitivity VLBI observations and novel calibration and imaging algorithms \citep[e.g.][]{Kim2025}. Utilizing multi-view phase-referencing observations \citep{Rioja2020} at 15 GHz, the centroid of the radio core can be measured with a precision of a few $\mu$as and monitored throughout the jet formation process. Such detailed observation of innermost jet activity over periods ranging from days to years is not possible with other AGNs, as they exhibit quite continuous jets \citep[e.g. M87$^{*}$,][]{Kim2025} or have black hole shadows with very small angular sizes \citep[e.g.][]{Nair2024}.

\textbf{Measuring the distance to nearby galaxies.} Measuring the distances to galaxies could be another way to define the cosmic distance ladder and to constraint the Hubble constant. Few 6.7~GHz CH$_3$OH masers have been detected in star-forming regions (or clusters) within external galaxies (e.g., the Large Magellanic Cloud and M31), with peak luminosities ranging from $\sim$ 500 to $2.3\times10^4$~Jy~kpc$^2$ \citep{Chen+2022}. For a galaxy 500 kpc away, the expected maser flux density would be $\sim$ 2–92 mJy, enabling corresponding astrometric precision better than $\sim$20 $\mu$as using 2-hr integration and a bandwidth of $\sim$200 kHz with phase-reference observations. This precision would be sufficient to resolve linear motion of $\sim$100 km s$^{-1}$ within a galaxy, thereby facilitating distance measurements \citep[][where the tracer is water maser at $\sim$22 GHz]{Brunthaler+2008}. The first candidate used to test the feasibility of the method could be the Large Magellanic Cloud. Furthermore, the JRT is well-suited for searching for such 6.7~GHz CH$_3$OH masers in external star-forming galaxies. 
Table~\ref{tab:fringe_ska_jrt_combined} explicitly lists the SKA-Mid--JRT fringe limits for the relevant SKA-Mid bands.

\section{Summary}

The Jingdong Radio Telescope (JRT) is a 120~m fully steerable single-dish facility with broad frequency coverage and high sensitivity. Its relatively southern location ensures wide sky access, including the Galactic Center and southern pulsar populations that are inaccessible to many northern-hemisphere telescopes.

We introduced the location, antenna and system design, and major single-dish scientific goals of JRT. Then we discussed the synergy between SKA-Mid and JRT, the two facilities share long common view time ($\sim$4 h) for equatorial sources (Dec.:$-45^{\circ}$ to $25^{\circ}$ ) and form highly sensitive intercontinental baselines, substantially improving $uv$-coverage and imaging sensitivity for existing VLBI networks. Our fringe-detection analysis shows that the SKA-Mid--JRT baseline achieves sub-mJy single-baseline detection limits across the SKA-Mid observing bands, providing a lower fringe detection threshold $S_{\rm fd}$ and improving the detectability of faint compact emission, especially on the outer edge of the $uv$-coverage. This joint capability would enable $\mu$as-level astrometry of PSR~J0437$-$4715 to provide the most stringent constraint on $\dot{G}/G$ for fundamental physics; high–sensitivity VLBI observations of M60$^{*}$ would reveal the innermost jet formation in horizon-scale spatial resolution; high–precision VLBI astrometry of 6.7\,GHz CH$_3$OH masers in nearby star-forming galaxies, starting with the Large Magellanic Cloud, would enable geometric distance measurements at the $\sim$20\,$\mu$as to define the cosmic distance ladder and to constraint the Hubble constant.

Beyond pulsar timing and nanohertz gravitational wave search, JRT is expected to involving VLBI, ensuring strong contributions to international networks. The synergy between JRT and SKA-Mid would jointly enhance the VLBI capabilities into a new era of study of faint sources and precise astrometry.

\section*{Acknowledgements}
W. Chen is supported by the National Natural Science Foundation of China (NSFC) under grant No. 12573073, Yunnan Fundamental Research Projects (grant No. 202401AT070144) and Yunnan Foreign Talent Introduction Program (grant No. 202505AO120021); Z. Li is supported by the NSFC under grant No. 12173087; N. Liu is supported by the NSFC under grant No. 12573070 and 12373074.

\bibliographystyle{abbrvnat-maxbibnames4}
\bibliography{chapter} 

\end{document}